\documentclass[review,3p,twocolumn]{elsarticle}

\usepackage{lineno,hyperref}

\usepackage[utf8]{inputenc}    
\usepackage[T1]{fontenc} 
\usepackage[english]{babel} 
\usepackage{graphicx}
\usepackage{enumitem} 
\usepackage{hyperref} 
\usepackage{upgreek}
\usepackage{siunitx}
\usepackage{xcolor}
\usepackage{float}

\journal{}

\begin{document}

\begin{frontmatter}
  \title{Characterization of a triple-GEM position sensitive detector for X-ray fluorescence imaging} 
  \author{Geovane G. A. de Souza\corref{corauthor}}
  \cortext[corauthor]{Corresponding author}
  \ead{geovane.souza@usp.br}
  \author{Hugo Natal da Luz}
  \address{Instituto de F\'isica, Universidade de S\~ao Paulo\\Rua do Mat\~{a}o 1371, 05508-090 Cidade Universit\'{a}ria, S\~{a}o Paulo, Brasil}
\begin{abstract}
In this work we characterized a X-ray position sensitive gaseous detector based in a triple stack of gas electron multipliers (GEM). The readout circuit is divided in 256 strips for each dimension and using a resistive chain interconnecting the strips, we are able to reconstruct the radiation interaction points by resistive charge division. The detector achieved gains above $10^4$, energy resolution of 15.28\,\% (FWHM) for 5.9\,keV X-rays, and position resolution of 1.2\,mm, while operating in Ar/CO$_2$(90/10) at atmospheric pressure. 

\end{abstract}

\end{frontmatter}


\section{Introduction}

X-ray fluorescence is an important technique on elemental analysis whenever a non-invasive and non-destructive method is required. New detectors that are able not only to identify material composition, but also their spatial distribution, improve this technique that can be applied on cultural heritage studies, archaeology and geology. This requires the ability to determine the position of interaction, while measuring the X-ray energy deposited in the detector.

The Gas Electron Multiplier~(GEM) is a proportional counter, introduced in 1997~\cite{Sau97,Sau16}. This Micropattern Gaseous Detector~(MPGD) is composed of \SI{50}{\micro\meter} thick kapton foil coated on both sides with \SI{5}{\micro\meter} copper layers. The foil is perforated with bi-conical holes (\SI{50}{\micro\meter} and \SI{70}{\micro\meter} diameter in the kapton and in the copper, respectively) etched in a hexagonal pattern. By applying an appropriate potential difference between the two sides of the GEM, a very large electric field is created inside the holes. This results in a focusing of the electric field lines towards the holes, provided that a weak uniform electric field is defined above the top electrode. With the GEM immersed in an adequate gas mixture, Townsend avalanches occur inside the holes, amplifying the charge created by a primary ionization in the gas.

With this type of detector, it is possible to build large detection areas with fair position and energy resolution even at high counting rates. These properties have included GEMs among the preferred choices for High Energy Physics experiments (for example~\cite{ALICEUP}--\cite{Alt02}). It is possible to use such technology not only to detect charged particles on these experiments but also $\gamma$ and X-ray radiation making these detectors good candidates for full-field XRF imaging \cite{Wro16}. It is well known that the energy resolution of gaseous detectors has limitations when compared to solid state detectors, due to the smaller number of electrons produced in each interaction, leading to higher statistical fluctuations in the primary cloud. Discussions about this are abundant in the literature and can be found, for example in~\cite{Knoll}. However, gaseous detectors become an interesting tool when large areas must be studied, without the need of sample scans thanks to the simultaneous measurement of the position of the radiation interaction and its energy over areas of hundreds of \si{\square\cm}.

\section{Experimental setup}

\begin{figure}[h]
\centering
\includegraphics[width=7.5cm]{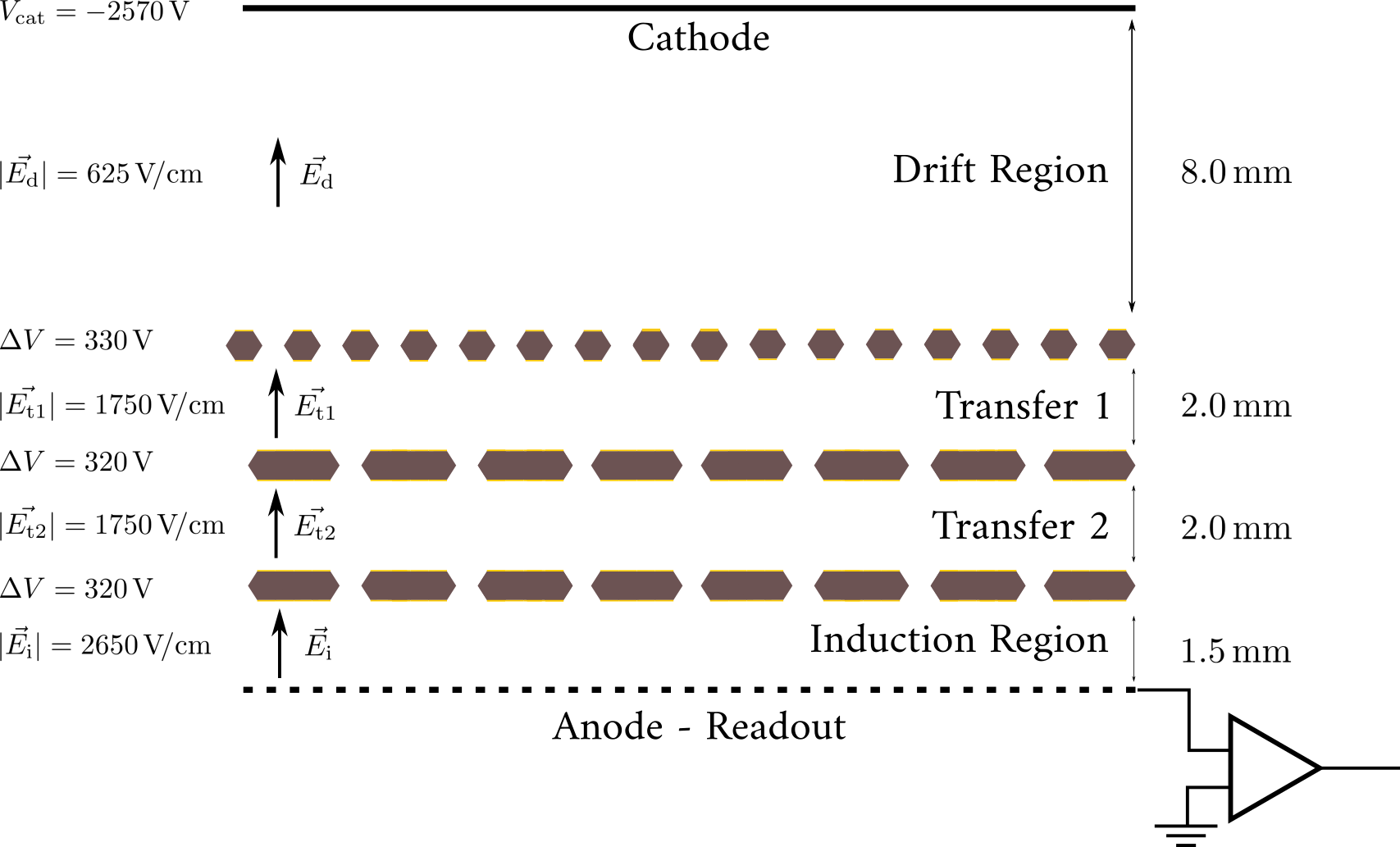}
\caption{Full detector setup}
\label{ImaGEM2}
\end{figure}

The detector consists on a cascade of GEMs immersed in a mixture of Ar/CO$_2$\,(90/10) at atmospheric pressure. Two different geometries were used in this study: a cascade of two GEM foils with a pitch of \SI{140}{\micro\meter} between the holes (this is the standard --- S --- pitch); and a cascade of three GEM foils, where the first GEM (the one on the top) had a pitch of \SI{90}{\micro\meter} (the small pitch --- SP --- GEM). The final triple-GEM geometry can be seen in figure~\ref{ImaGEM2}, where the dimensions and typical electric fields and voltages are also depicted. The total active area of the detector is $10\times 10$ cm$^2$.

\begin{figure}[h]
\centering
\includegraphics[width=7cm]{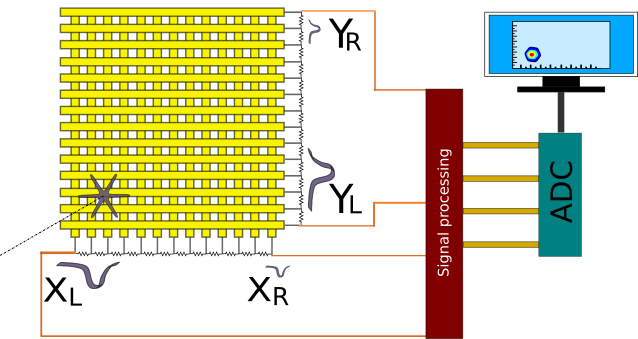}
\caption{Scheme of the segmented readout system using resistive chains.}
\label{readout}
\end{figure}

The detector window is a \SI{50}{\micro\meter} kapton foil and the cathode is the same type of foil, but coated with a \SI{5}{\micro\meter} thick copper layer. The space between the two surfaces is around 5 mm. These materials were used for this prototype and can easily be changed by others that will attenuate the intensity of lower energy X-rays by a smaller amount.
The readout system is segmented in 256 strips in each dimension~(fig.\ref{readout}), which are interconnected through resistive chains. The strips are \SI{100}{\micro\meter} wide with a pitch of \SI{400}{\micro\meter}. By collecting the the charge at both ends of each resistive chain, it is possible to calculate the projection of the primary X-ray ionization on the X--Y plane for each coordinate through a trivial `center of mass' algorithm (eq.~\ref{interaction}). 

\begin{equation}
    x= l \frac{X_L-X_R}{A},\qquad
    y = l \frac{Y_L-Y_R}{A}
    \label{interaction}
\end{equation}

\noindent where:
\begin{itemize}
\item $X_L$, $X_R$, $Y_L$ and $Y_R$ are the signal amplitudes for the
  left and right ends of the $X$ and $Y$ resistive chains according to
  figure~\ref{readout}; 
\item $l$ is the length/width of the detector and
\item $A$ is given either by the sum of the amplitudes of all four
  channels, or by the amplitude of the signal collected from the
  bottom electrode of the last GEM.
\end{itemize}
The signal from the bottom electrode of the GEM also served as the global trigger of the electronic system. 

The resistive chains were composed of SMD resistors on a printed circuit board designed to interconnect all the readout strips when plugged to the standard 128-pin connectors of the readout board. The charge collected from each electronic channel is integrated by a standard charge sensitive pre-amplifier and shaped by a shaping amplifier. After application of simple logic, it is sampled by a 12 bit ADC. All the electronic devices are standard off-the-shelf nuclear instrumentation modules.

As X-ray source we are using the Amptek Mini-X with a silver target, operating at a high voltage of typically 15\,kV and current around 15\,$\upmu$A. This results in an energy distribution in the detector between around \SI{2}{\kilo\electronvolt} and \SI{15}{\kilo\electronvolt}. The low energy limit is due to the detector window and cathode, which result in \SI{100}{\micro\meter} thickness kapton and to the amount of air between the X-ray source and the detector. At the higher energies, the absorption efficiency of the \SI{8}{\milli\meter} thick absorption layer makes the detected spectrum drop sharply for higher energies. The X-ray spectrum as collected in the detector has a maximum intensity at around \SI{7}{\kilo\electronvolt}. We are also using a $^{55}$Fe radioactive source, which decays into manganese by electron capture emitting 5.9\,keV (K$_{\alpha}$) and 6.4\,keV (K$_{\beta}$) characteristic X-rays (with relative probabilities of 100 and 20, respectively)~\cite{Pen72} for energy calibration and to calculate energy resolution. In general, the counting rates for these measurements were kept at around 800 Hz. The counting rate of the detector is not limited by the GEM, which can stand rates of the order of several MHz per \si{\square\cm}~\cite{Sau16} (and references therein), but by the resistor chain, which does not allow the collection of two different simultaneous hits in the detector surface. Taking into account a shaping time of \SI{1}{\micro\s} in the amplifier for a correct determination of the pulse amplitude over the whole area of the detector, the estimated maximum counting rate is around \SI{10}{\kilo\hertz\per\cm\squared}.

A framework for data processing, image reconstruction and analysis was developed using ROOT~\cite{ROOT} and other C++ libraries.


\section{Results}

\begin{figure}[h]
\centering
\includegraphics[width=7cm]{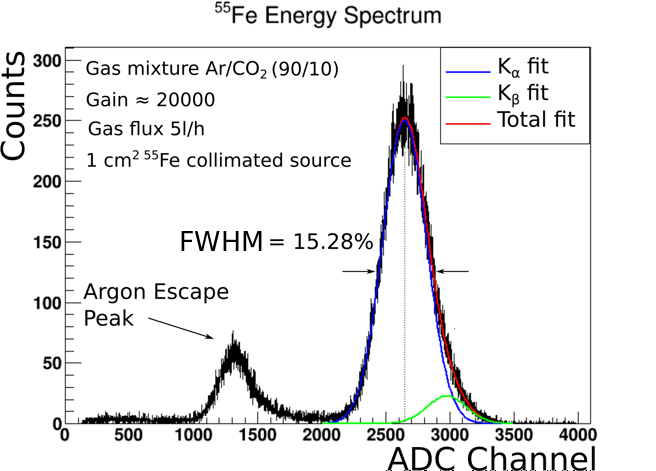}
\caption{Energy spectrum for the iron radioactive source, $^{55}$Fe. The spectrum was obtained irradiating a small area of the detector, around \SI{1}{\centi\meter\square}$^2$}
\label{energyres}
\end{figure}

The detector energy calibration and characterization in terms of energy resolution has been done using the $^{55}$Fe source. The energy resolution was optimized by tuning the drift and transfer fields of the GEM stack, achieving 15.28\% full width at half maximum (FWHM) for the energy of 5.9\,keV, irradiating an area of around \SI{1}{\square\cm} of the detector, working at a gain in charge above $10^4$ (fig.~\ref{energyres}), optimizing the position and energy resolution. Because of the resistive division and the electronics used, it is expected that for high X-ray rates, effects of pile-up will impair the energy resolution. To avoid this, the counting rate was kept low, allowing the correct measurement of the intrinsic energy resolution of the detector. The main peak is a convolution of the two manganese K-lines and the peak around 3\,keV is the argon escape peak, resulting from fluorescence X-rays escaping the detector absorption region. GEM detectors have slight gain variations throughout their active area, degrading the energy resolution when the whole area is irradiated. There are plans to apply adequate gain corrections that will allow to reconstruct the energy spectrum, such as shown in other works (see for example~\cite{Vel10}). This spectrum is not influenced by the resistive charge division because it is collected from the bottom electrode of the last GEM, completely decoupled from the resistor chains.

\begin{figure}[h]
	\centering
	\includegraphics[width=7cm]{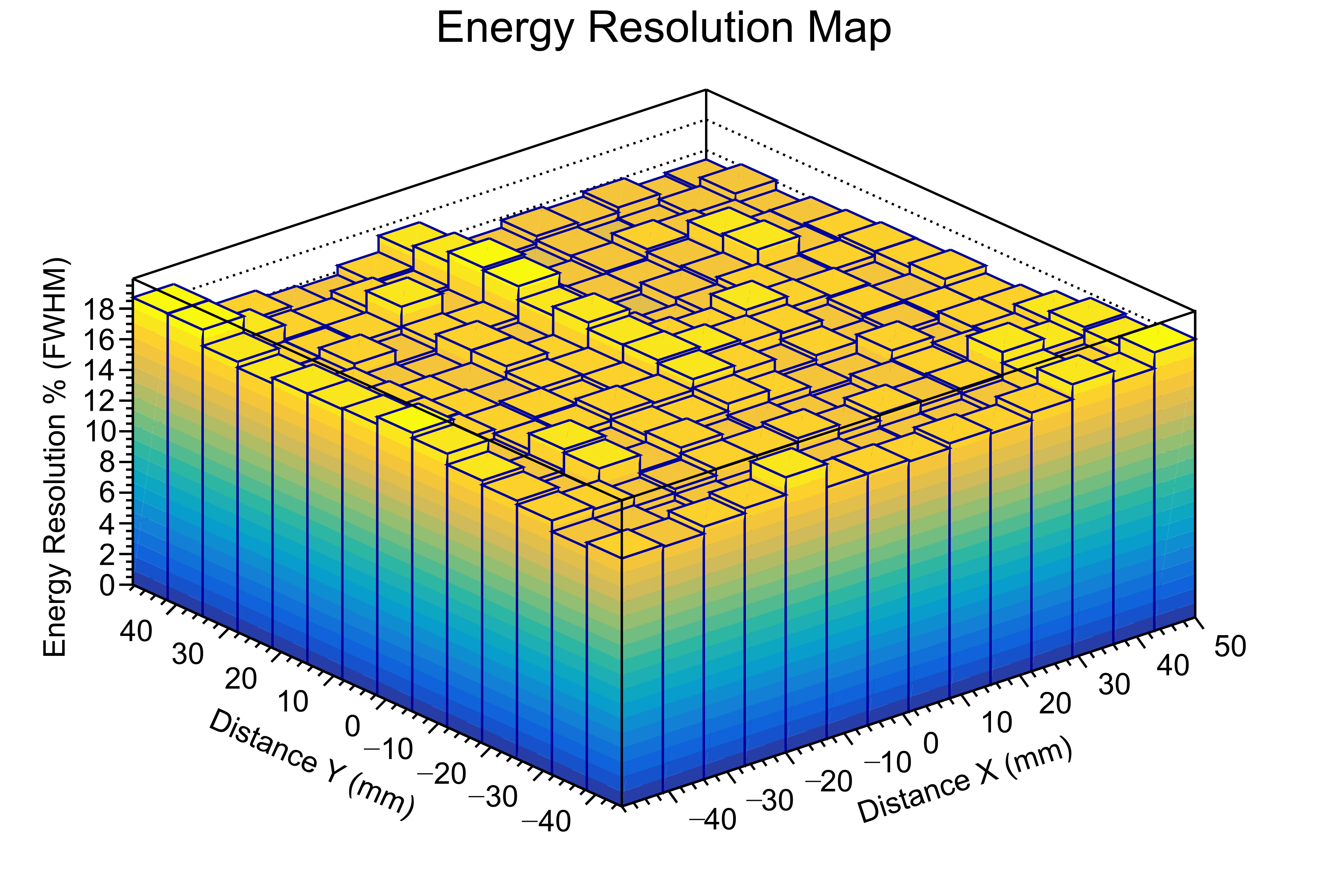}
	\caption{Energy resolution for each region of the detector measured using $^{55}$Fe radioactive source.}
	\label{mapares}
\end{figure}

\begin{figure}[h]
	\centering
	\includegraphics[width=7cm]{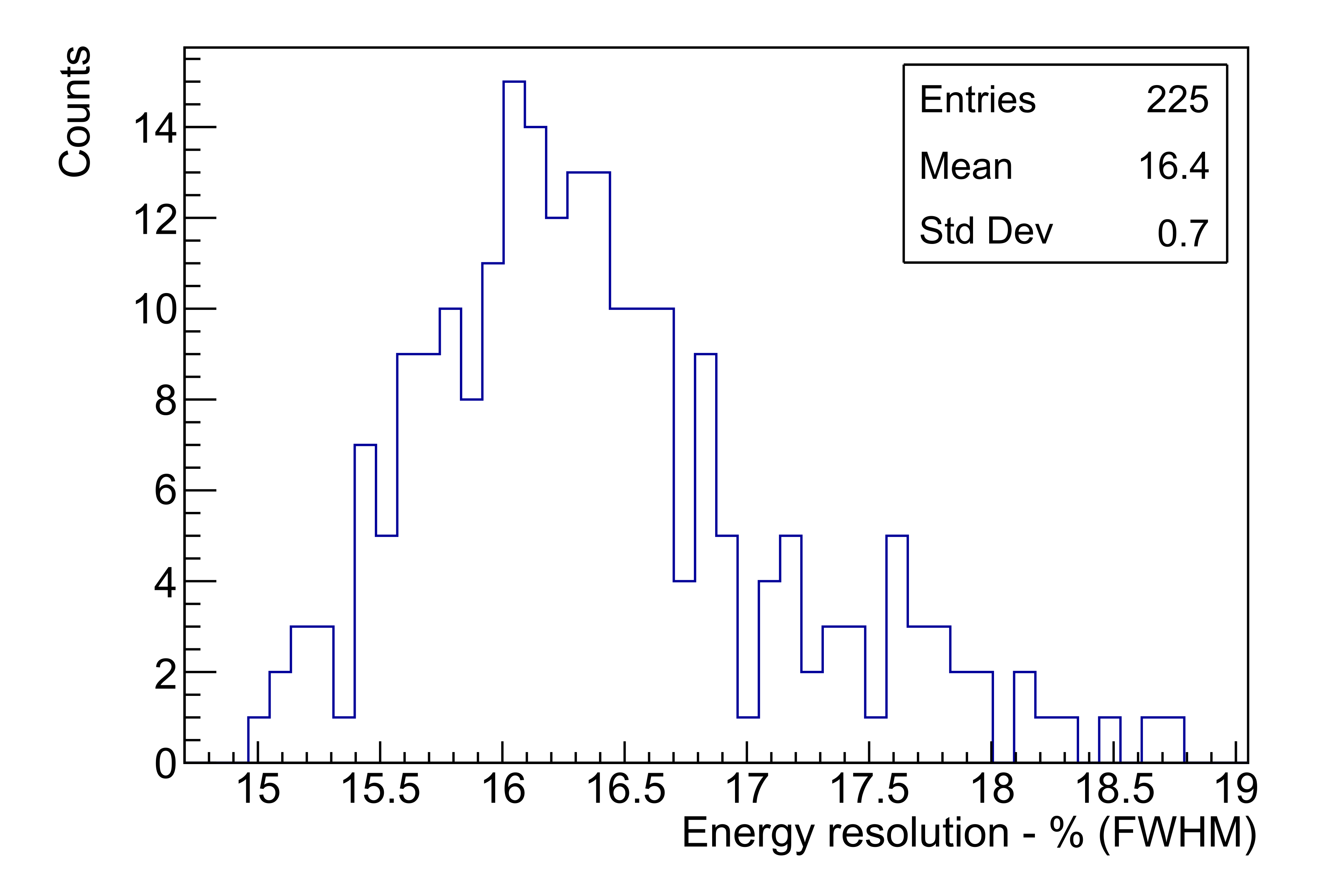}
	\caption{Energy resolution distribution for the 225 regions.}
	\label{hist_de_resolucao}
\end{figure}

To study the deviation of the energy resolution across the detector's effective area, the whole detector was irradiated using the $^{55}$Fe source and in the analyses, the total area was divided into 225 squares. For each one of these regions, the energy resolution was determined. The distribution of the energy resolution throughout the whole sensitive area of the detector can be seen in figures~\ref{mapares} and~\ref{hist_de_resolucao}. The energy resolution has a standard deviation of 0.7\% for all the 225 regions, with the worst value below 19\% 

As an imaging system, the detector was characterized by measuring its position resolution and contrast. The position resolution is defined by the width of its Point Spread Function (PSF) which would be the width of the image of an infinitely small point. If one is considering only one dimension, the Line Spread Function (LSF --- the width of the image of an infinitely thin slit) can also be considered. Due to the practical impossibility of imaging such small objects, other more realistic approaches can be used to derive these functions. References~\cite{medicalimage,MedImg1} contain detailed descriptions of these concepts for characterization of imaging systems.

\begin{figure}[h]
\centering
\includegraphics[width=7cm]{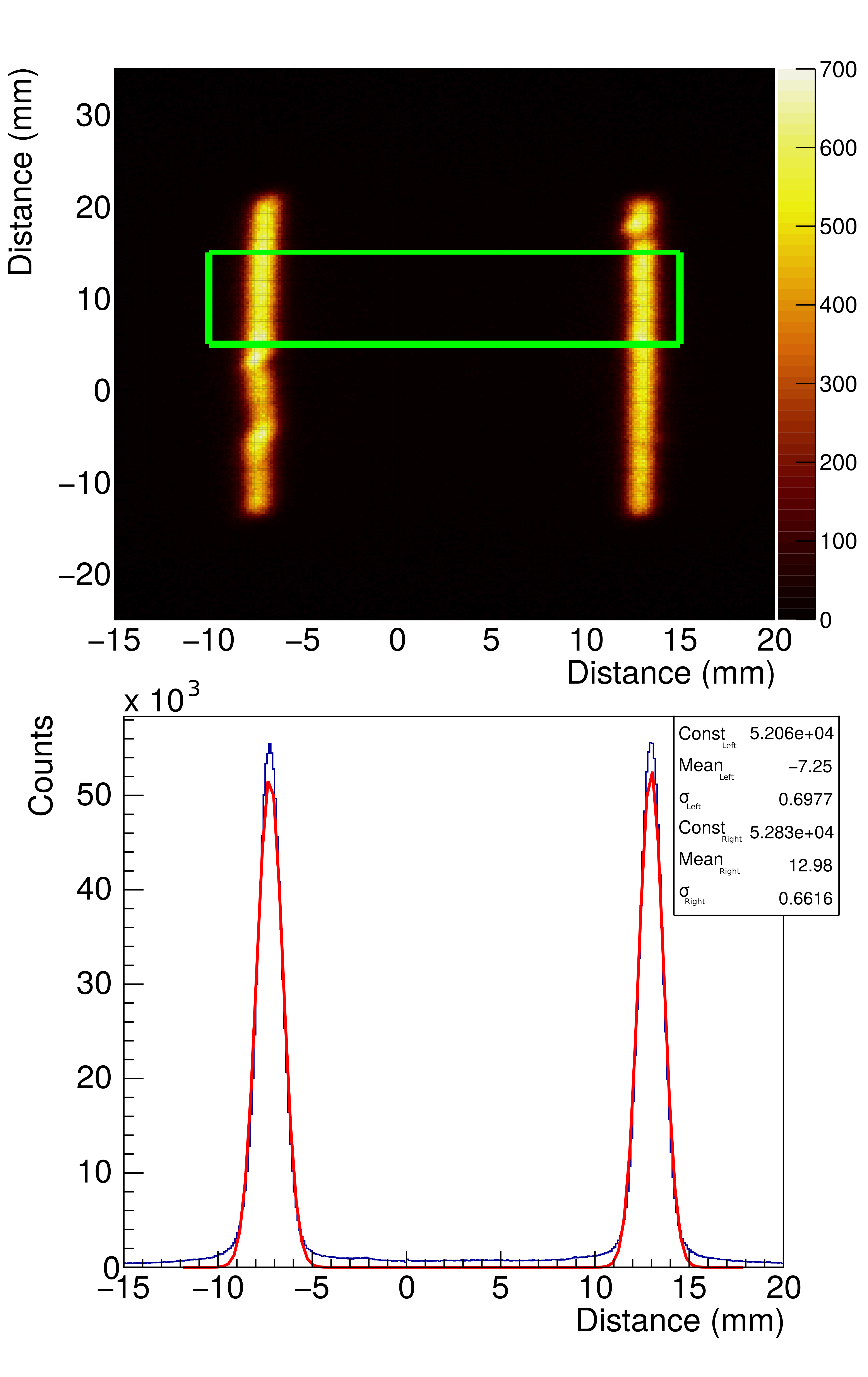}
\caption{Double slits method. The area defined by the green rectangle is the selected data to create the profile shown below.}
\label{fendasduplas}
\end{figure}

Three different methods were applied. In the first one, a 1\,mm thick stainless steel plate with two thin slits was placed directly above the detector's window. If the width of the slits is slightly smaller than the position resolution, the width of their image will give an approximate idea of the width of the LSF. The final result was above 1\,mm, as will be shown and described below. Therefore, the width of the slits chosen was 1\,mm. They were separated by 20\,mm to allow for position calibration, and conversion between pixels and length units, and consequent determination of the width of their image. The detector was irradiated for one hour with the X-ray tube. To calculate the spatial resolution, two Gaussian curves were fitted, one to each slit profile (fig. \ref{fendasduplas}). The resolution is the average of the full width at half maximum of both fits. The spatial resolution obtained by this method using the whole energy spectrum, collected from the Amptek Mini-X source, is 1.79\,mm, as shown in figure~\ref{fendasduplas}. Position resolution measurements were also performed for the $y$-direction. The position resolution was slightly worse and can be due to noisier electronic channels, defective resistive chains or problems in the readout. There are no reasons to believe that the triple-GEM setup would have a noticeable difference in performance for the different coordinates. The position resolution as a function of the energy for different setups is discussed in section~\ref{sec:disc}.

Although this double slit method is good to measure the spatial resolution, it does not give any information of the image contrast. The contrast discrimination is important to evaluate how well an imaging system can distinguish differences in luminosity~\cite{medicalimage}. In order to measure both contrast and resolution for the same image, we used two other methods.

The second one was the analysis of an image of a sharp edge placed above the detector window. The image obtained is the Edge Spread Function~(ESF) of the detector. The profile of the image is a step function and by calculating its first derivative by numerical methods, we obtain the Line Spread Function~(LSF), which is an indicator of the detector resolution in one dimension, as explained before. 

\begin{figure}[h]
\centering
\includegraphics[width=7cm]{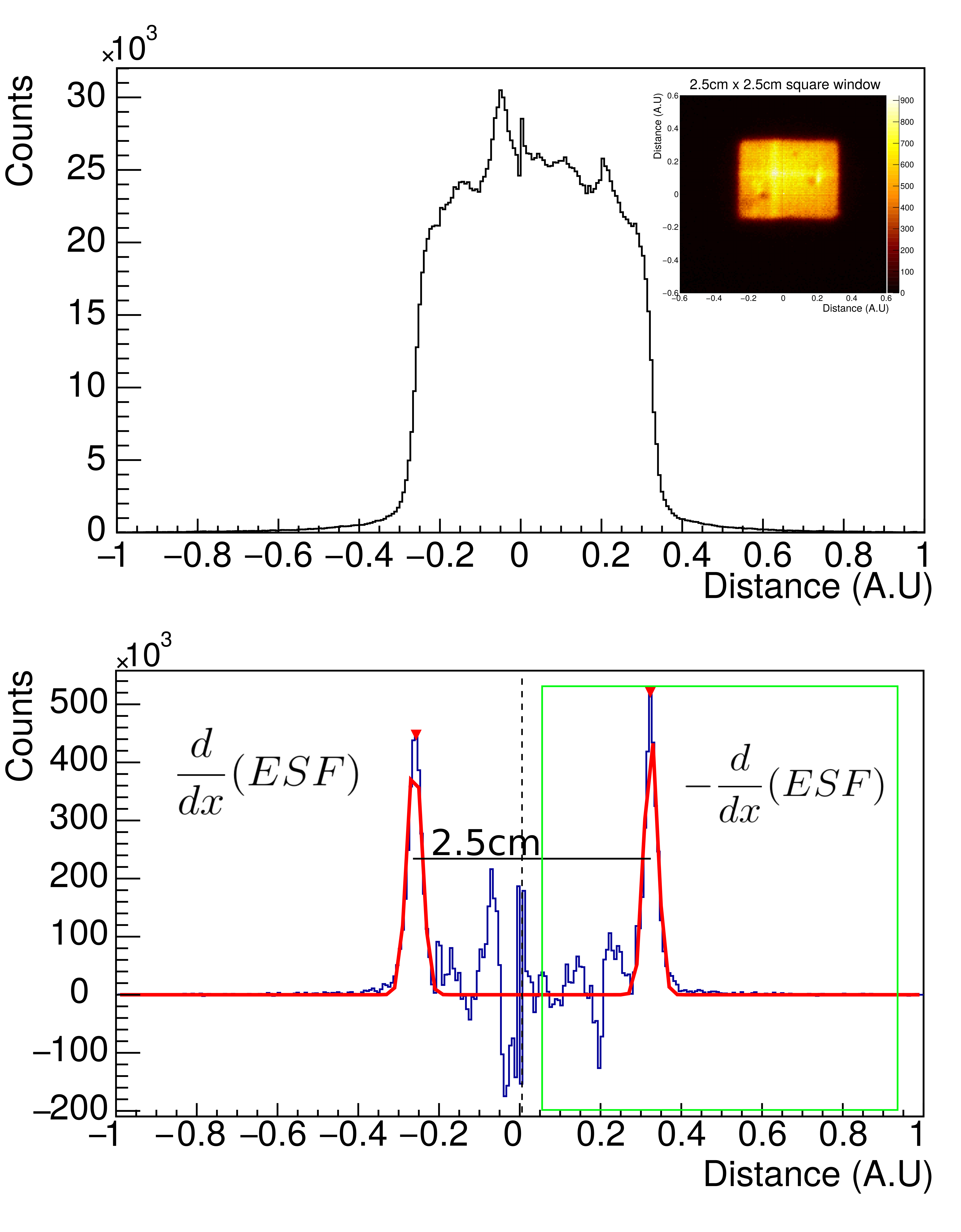}
\caption{The profile of a $2.5\times 2.5\,$\si{\square\cm} window for both dimension axis. The red square indicates the selected data.}
\label{ESF}
\end{figure}

\begin{figure}[h]
\centering
\includegraphics[width=7cm]{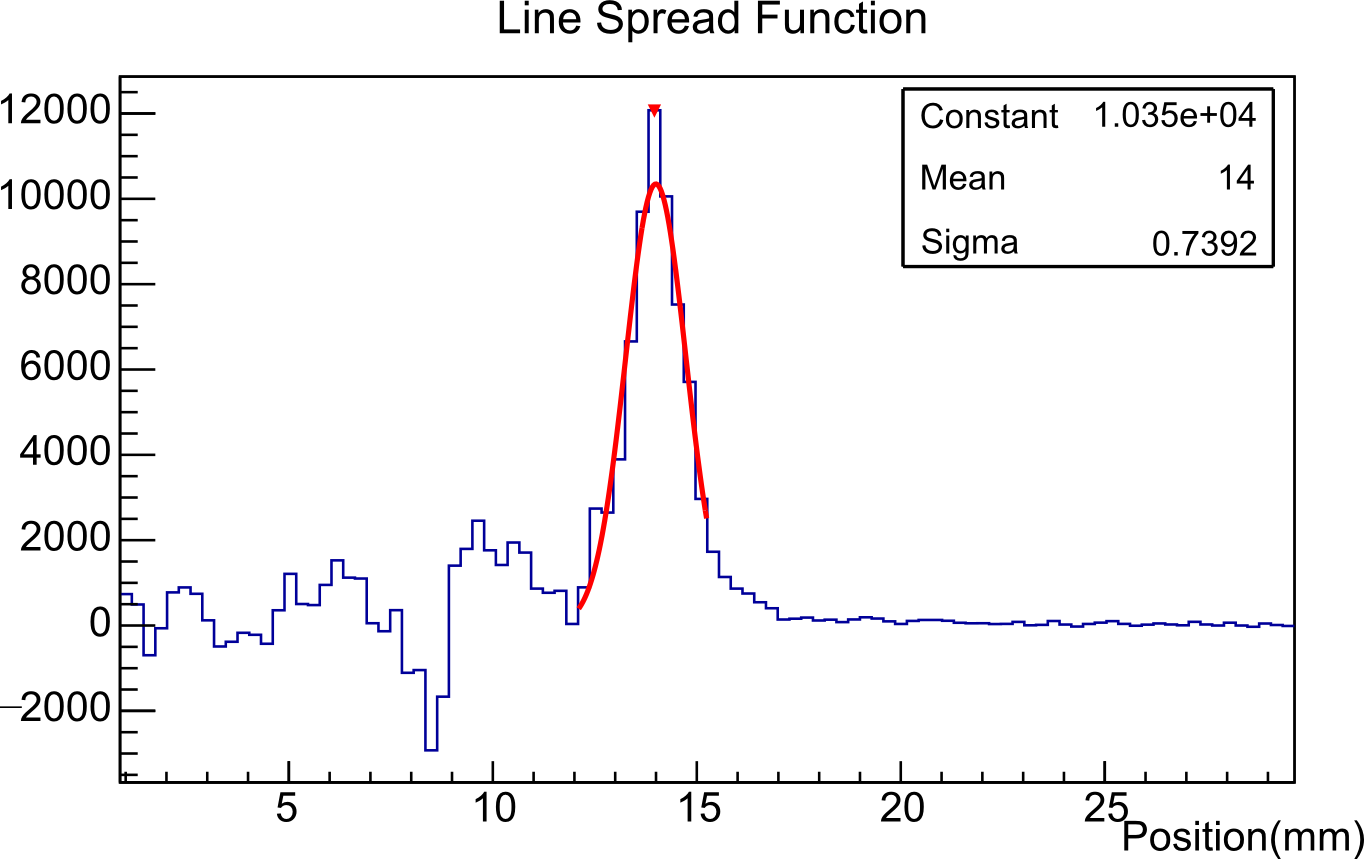}
\caption{The derivative of the ESF indicates the spatial resolution.}
\label{LSF}
\end{figure}

To use this method, a square opening of $2.5\times2.5\,\si{\square\cm}$ was imaged. The two edges at known distance were used to calibrate the image, as seen in fig.~\ref{ESF} and the one inside of the green rectangle was used as the ESF as shown in fig.~\ref{LSF}.

The different intensities in the image of fig.~\ref{ESF} are reflected in the X-profile. The darker regions in the image are due to small defects in the foils caused by sparks that create less sensitive zones and the higher intensity regions can be related to non-uniformities in the construction of the resistive chains. These two effects can easily be mitigated in future prototypes.

This measurement has the advantage of exploiting the range of spacial frequencies of the detector. For a realistic spatial resolution, even if two objects very close to each other are distinguishable, they have less contrast. A smaller decrease in the contrast for objects very close to each other means a better imaging system. The Fourier transform of the LSF results in the contrast as a function of the spacial frequency. This is defined as the Modulation Transfer Function~(MTF). Using the MTF we are able to evaluate the variation of the contrast with the spatial frequency, which is related to the position resolution~\cite{medicalimage,MedImg1}.

\begin{figure}[h]
\centering
\includegraphics[width=7cm]{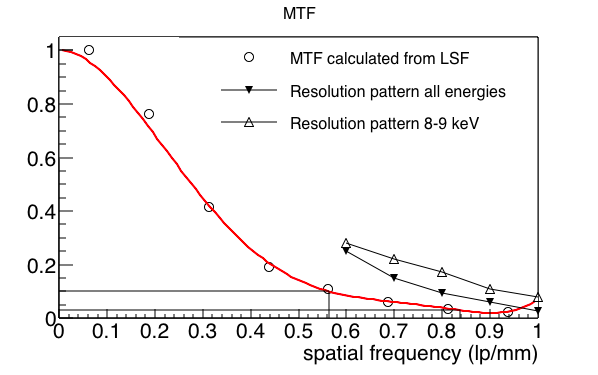}
\caption{The circles represent the Fourier coefficients of the LSF. The black triangles are contrast points calculated using the resolution pattern of fig.~\ref{contrast} for the shown X-ray image.}
\label{MTF}
\end{figure}

Figure~\ref{MTF} shows the MTF resulting from the Fourier Transform of the LSF from fig.~\ref{LSF} (black circles), where the red curve serves to guide the eye. The contrast drops as expected for higher frequencies. The frequencie at contrast of 10\,\%  is marked in the figure and the value obtained is 0.56\,lp/mm (line pairs per millimeter), which is consistent with the width of the LSF and the width of the image of the slits from fig.\,\ref{fendasduplas}.

\begin{figure}[h]
\centering
\includegraphics[width=7cm]{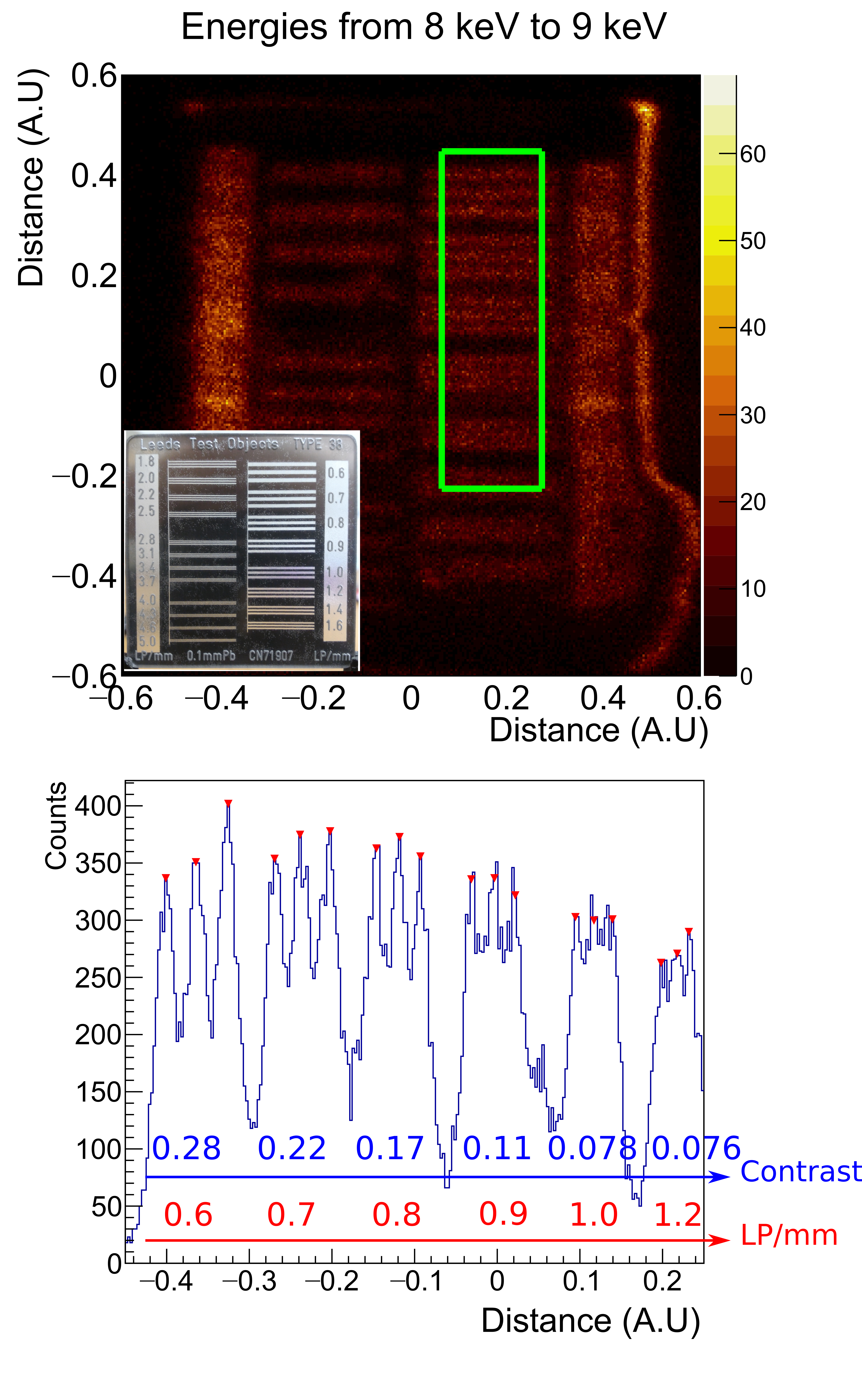}
\caption{Above: Resolution pattern used to measure the contrast. The area limited by the green rectangle is the data selected for the analysis. Below: The intensity profile of the green rectangle marked in the image. The contrast is calculated by the Michelson contrast definition \cite{michelson}}
\label{contrast}
\end{figure}

The third method measures directly both contrast and resolution using a resolution pattern. The pattern consists of sets of slits with specific widths and at specific distances, corresponding to specific spatial frequencies. The difference between illuminated and dark regions is related to the contrast for each spacial frequency~(fig.~\ref{contrast}). 

In figure~\ref{MTF}, the normalized contrast measured directly from the resolution pattern was also plotted for the total energy spectrum and the energy interval from 8 to 9\,keV. The contrast in each group of slits was determined by subtracting the average of the valleys from the average of the peaks and dividing by their sum, according to Michelson~\cite{michelson}. The determination of the MTF from the edge spread function seems to underestimate the performance of the detector for larger objects. The contrast for the energy range from 8 to 9\,keV is higher because it is the optimal range of the detector (see discussion below).

\section{Discussion}
\label{sec:disc}

The performance of this type of detector as imaging system, when using the resistive charge division varies throughout the energy spectrum. There are two factors that influence the position resolution: the signal-to-noise ratio (SNR) and the photo-electron range. While the SNR becomes more important for the lower energies due to the smaller amplitude of the signals, for higher energies the range of the photo-electrons ejected from the argon atoms increases. Since the resistive charge division determines the center of mass of the primary cloud, which is shifted from the point of interaction, its size increases the uncertainty.

\begin{figure}[h]
\centering
\includegraphics[width=7cm]{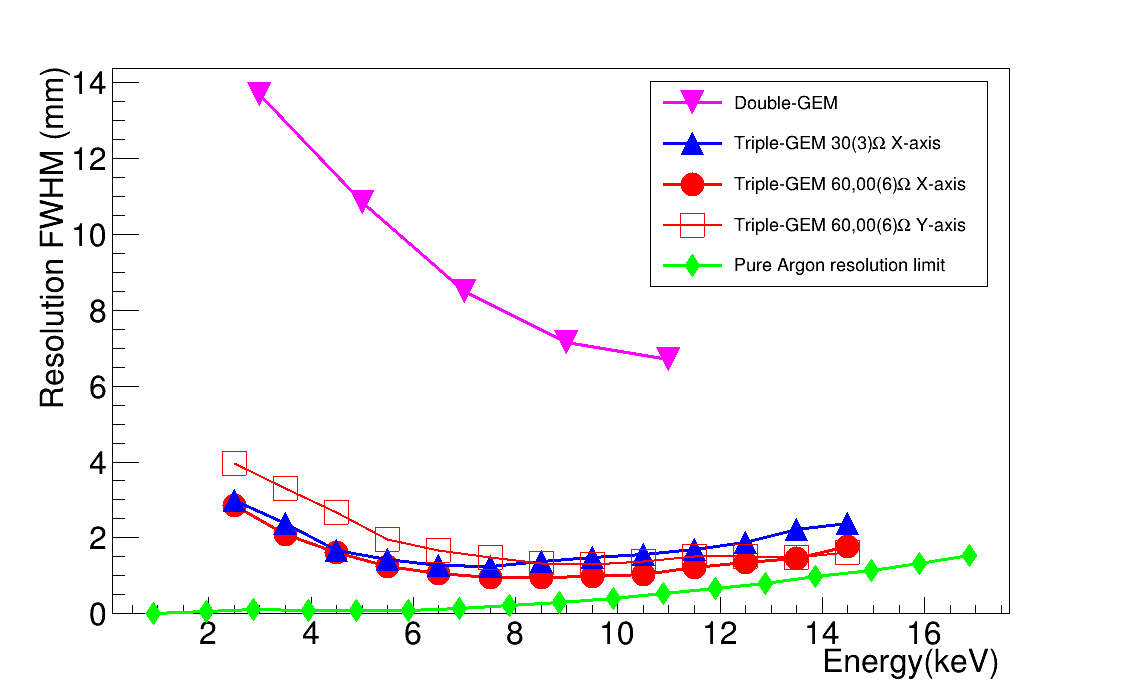}
\caption{Spatial resolution as a function of the photon energy. The green line indicates the pure Argon resolution limit as simulated in~\cite{Aze15}.}
\label{resolution}
\end{figure}

Figure~\ref{resolution} shows the position resolution achieved with the method of the two slits for different X-ray energies. The different colors correspond to different experimental setups. The inverted magenta triangles represent the data for a double-GEM system, with a poorer resolution due to the impossibility to reach very high charge gains. The blue triangles and the red circles represent triple-GEM assembly using resistive lines with $30(3)\,\Omega$ resistors between each strip and $60.00(6)\,\Omega$, respectively  (note the precision of 10\,\% and 0.1\,\% of the values). From literature \cite{Cus02} it is known that for systems using charge division principle on strips, increasing the resistance value and decreasing resistance deviations between resistors in the same chain, the position resolution can be improved. This plot shows a slight improvement for energies above 6\,keV. For comparison, published simulated data of the resolution limit due to the photoelectron range expected for pure argon in a 10\,mm thick detector is also plotted~\cite{Aze15} (green squares). The difference for a thickness of 8\,mm is very small given the photoelectron range at these energies, which is much smaller than the detector. The position resolution of the $y$-coordinate is also shown. It is poorer due to the higher noise in these electronic channels, as explained before.

For energies above 6\,keV the resolution curve is very close to the expected limit. By selecting different energy ranges the detector can improve the position resolution with respect to the one obtained when the whole energy spectrum is used. The spatial resolution achieved for Ar/CO$_2$\,(90/10) was 1.2\,mm at the range of 8 to 9 keV, consistent with the points measured with the resolution pattern at this energy range and plotted in fig.~\ref{MTF}.

\begin{figure}[h]
\centering
\includegraphics[width=7cm]{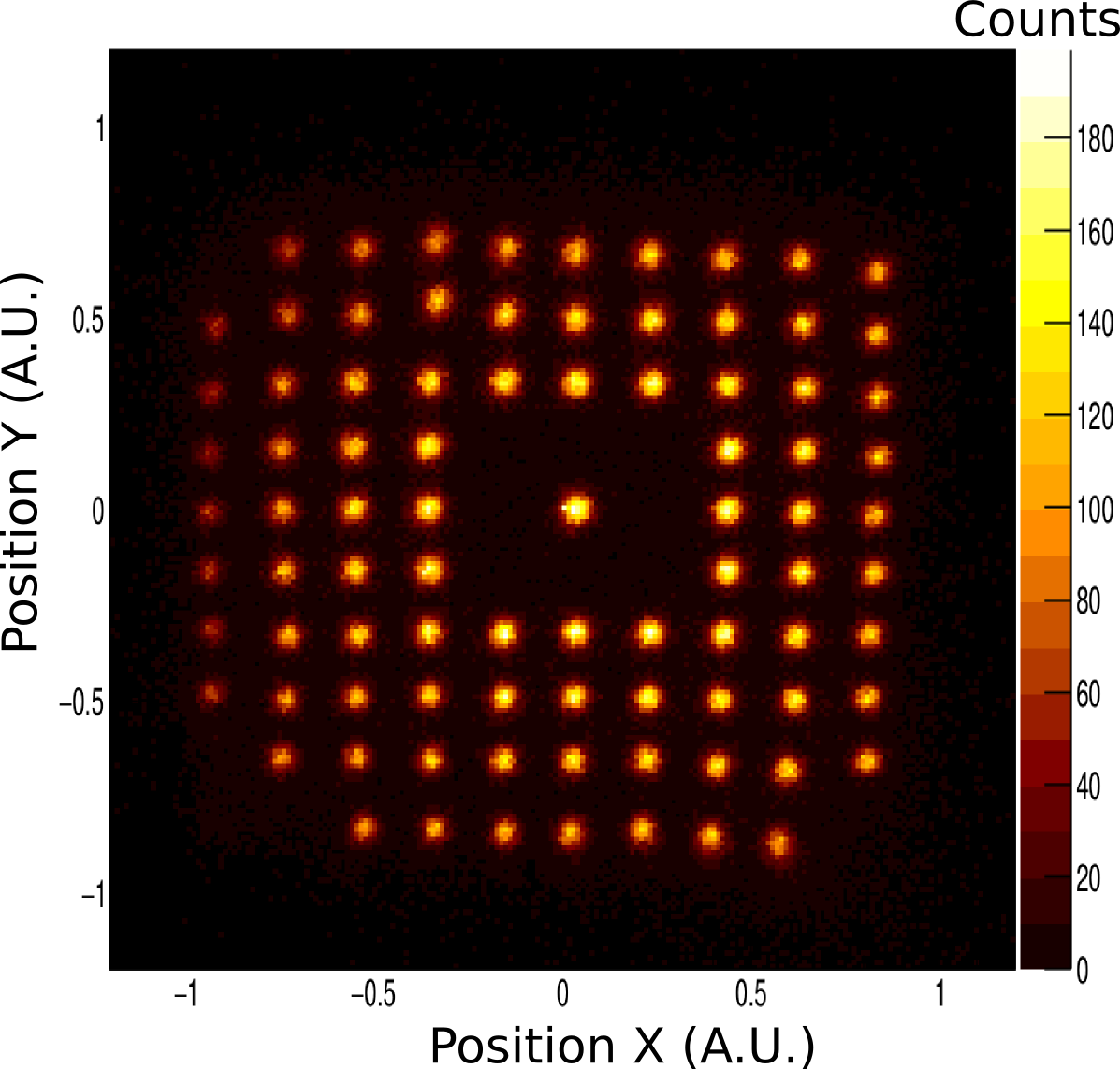}
\caption{Image of an array of holes drilled in a stainless steel plate. The holes have a pitch of 10\,mm and span most of the area of the detector..}
\label{fig:furos}
\end{figure}

The position resolution of the detector was measured mostly in central areas of the detector. Figure~\ref{fig:furos} shows the image of an array of 1\,mm holes drilled with a pitch of 1\,cm in a 1\,mm thick stainless steel plate, spanning a great part of its active area. The X-rays tube was places 1.5\,m away from the detector window and a 1\,mm collimator was used, resulting in an image that was more intense in the center than at the borders. It gives a qualitative idea of the behaviour of the detector also near the edges, where the shape of the holes does not change significantly.

\section{Conclusion and Future Work}

A triple-GEM X-ray detector has been tested for imaging purposes with resistive charge division using five electronic channels. It achieves a gain in charge well above $10^4$, which results in an energy resolution of 15.28\,\% for 5.9\,keV X-rays and position resolution around 1.2\,mm for energies between 8 and 9\,keV. 

By increasing the detector gain in charge, working at higher potentials across the GEM plates, better position resolution could be achieved at lower energies, but this would also increase the discharge probability.

The system performance was described in transmission mode, but the plan is to make fluorescence imaging in the future using a pinhole. Deeper studies are in progress, namely the variations of the position resolution as a function of the position in the detector, the quantification of the image distortions due to imperfections of the resistive chains and the possibilities to correct for these effects.

The gain across the detector's sensitive area is not completely uniform. This is caused mainly due to irregularities on the GEM construction process and detector drift, transfer and induction regions. Therefore, improvements in the reconstruction and analysis framework are in progress to normalize and correct the gain over the whole sensitive area of the detector, improving the energy resolution. 

Future plans include the integration of the new ASIC SAMPA~\cite{SAMPA}, which was developed to work as a front end for the Time Projection Chamber and the Muon Chamber of the ALICE experiment at CERN after the upgrades for Run 3. The SAMPA chip is able to sample data from 32 different channels, which means that 8 chips will be enough to read discrete data from all the 512 readout strips, dramatically increasing the counting rate of the imaging system, while also improving the position resolution for lower energies.

\section*{Acknowledgments}

This work was supported by grants 2016/05282-2 and 2017/00426-9 from
Funda\c{c}\~ao de Amparo \`a Pesquisa do Estado de S\~ao Paulo, Brasil.



\end{document}